\documentclass[twocolumn,aps,amsfonts,amsmath,prd,nofootinbib]
{revtex4-1}
\usepackage[T1]{fontenc}
 \usepackage[latin1]{inputenc}
\usepackage{amssymb}
\usepackage{amsmath}
\usepackage{enumerate}

\def\be{\begin{equation}}
\def\ee{\end{equation}}
\def\ba{\begin{array}}
\def\ea{\end{array}}
\newcommand{\bea}{\begin{eqnarray}}
\newcommand{\eea}{\end{eqnarray}}
\def\vf{\varphi}
\def\Re{\mathop{\rm Re}\nolimits}
\def\Im{\mathop{\rm Im}\nolimits}
\usepackage{ifpdf}
\ifx\pdfoutput\undefined
   \pdffalse
   \usepackage{cite}
 \else
   \pdfoutput=1
   \pdftrue
  \usepackage[pdftex]{hyperref}
\fi

\makeatletter

\usepackage{epsf}

\def\be{\begin{equation}}
\def\ee{\end{equation}}
\def\ba{\begin{eqnarray}}
\def\ea{\end{eqnarray}}
\def\beas{\begin{eqnarray*}}
\def\eeas{\end{eqnarray*}}
\def\sla{\raise.15ex\hbox{$/$}\kern-.57em}

\newcommand{\SO}{\mathop{SO}}

\newcommand{\USp}{\mathop{{}USp}}
\newcommand{\OSp}{\mathop{{}OSp}}

\begin{document}

\title{\Large\bf Conjecture on hidden superconformal
\
symmetry of $N=4$ supergravity}

\author{
 {\bf Sergio Ferrara$^{1,2}$},
 {\bf Renata Kallosh$^3$},  and  {\bf Antoine Van Proeyen$^4$}}

\date{\today}

\affiliation{ $^1${\sl Physics Department, Theory Unit, CERN  CH
1211, Geneva 23, Switzerland }\\ $^2${\sl INFN - Laboratori
Nazionali di Frascati,  Via Enrico Fermi 40, 00044 Frascati,
Italy}\\  $^3${\sl Stanford Institute for
Theoretical Physics and Department of Physics, Stanford University, Stanford,
CA 94305}\\ $^4${\sl Instituut voor Theoretische Fysica, KU Leuven,\\
Celestijnenlaan 200D, B-3001 Leuven, Belgium}}

%\date{\today}

\

\

\begin{abstract}
 We argue that the observed UV finiteness of the 3-loop extended supergravities  may be a manifestation of a hidden local superconformal symmetry of supergravity.  We focus on the $SU(2,2|4)$ dimensionless
superconformal model. In Poincar{\'e} gauge where the compensators are fixed to $\phi^2= 6 M_P^2$
 this model becomes a pure classical $N=4$  Einstein supergravity.  We argue that in $N=4$ the higher-derivative superconformal invariants like $\phi^{-4}W^2 \bar W^2$
  and the consistent local anomaly $\delta(\ln \phi \, W^2)$ are not available. This conjecture on hidden local
  $N=4$ superconformal symmetry of Poincar{\'e} supergravity may be supported by subsequent loop
  computations.
\end{abstract}

\maketitle

\section{Introduction}

The purpose of this note is to address the following issue: what if extended
supergravity is perturbatively finite? Even if it is true (which of course
we do not know at present) why could it be important? Is it possible that the conjectured perturbative UV finiteness may reveal
some hidden symmetry of gravity? {\it Here we propose a conjecture that such
a hidden symmetry may be an $N=4$  local superconformal symmetry}. If the
4-loop $N=4$ supergravity is UV divergent, this conjecture will be
invalidated and, if it is UV finite, the conjecture will be supported.

Starting from the early days of supergravity, the superconformal calculus was
a major tool for constructing new Poincar{\'e} supergravity models, see for
example \cite{Ferrara:1977ij} and the   recent book \cite{TheBook}, which
describes in detail the superconformal origin of $N=1, 2$ supergravities,
including the role of the compensators and the gauge-fixing of the
superconformal models down to super Poincar{\'e}. Extended $N\leq 4$ supergravity
models were developed in Refs.
\cite{Bergshoeff:1980sw,Bergshoeff:1980is,deRoo:1984gd,deRoo:1985jh},
starting with superconformal symmetry.

$N>  4$ supergravity models  do not have an underlying superconformal symmetry.
This is related to the fact that  there are no matter multiplets, only pure supergravity multiplets are available.
In particular, there is no supersymmetric extension of the square of the Weyl tensor in $N>4$, as shown in Ref. \cite{deWit:1978pd}.

Here we would like to suggest a possibility that the superconformally
symmetric model underlying $N=4$ supergravity is not just a tool, but a major
feature of a consistent perturbative supersymmetric theory involving gravity.
Namely, we will show that the 3-loop finiteness of pure\footnote{Our analysis
is not valid for the case of $N=4$ supergravity interacting with matter,
studied in Ref. \cite{Tourkine:2012ip}. These models have a 1-loop UV divergence
\cite{Fischler:1979yk}.} $N=4$ supergravity \cite{Bern:2012cd}, taken together
with the absence of a candidate for a consistent local $N=4$ superconformal
anomalies suggests that the principle of local superconformal symmetry may
control the quantum properties of the gravitational theory, in the same way
as the principle of non-abelian gauge symmetry controls the quantum
properties of the standard model.

We are not used to thinking of a local four-dimensional conformal symmetry as a
reliable gauge symmetry, where the gauge-fixing and the ghosts structure
support the BRST symmetry  and the computations confirm the formal properties
of the path integral. The common expectation is that this symmetry may be
unreliable because of anomalies. Therefore we cannot use it for investigation of divergences in the usual Einstein gravity.

The $N=4$ local superconformal symmetry may be
an example of an anomaly-free theory, and therefore it is tempting to study
possible implications of the local superconformal symmetry, starting with
this case. In particular, the absence of the 3-loop UV divergences in pure
$N=4$ supergravity \cite{Bern:2012cd} may be interpreted as a manifestation of
the superconformal symmetry of the un-gauge-fixed version of this theory.

We will discuss possible implications of the conjecture of hidden
$N=4$ superconformal symmetry for the all-loop UV properties of $N\geq 4$
supergravities \footnote{Some early hints about the possibility of UV
finiteness of $N\geq 5$ and not only $N=8$ supergravity were given in
Ref. \cite{Bern:2007xj} based on the observation
 that generic theories of quantum gravity based on the Einstein-Hilbert action may be better behaved in UV at higher loops than suggested by naive power
 counting.}. In $N>4$, in absence of duality anomalies \cite{mar}, the duality
current conservation argument can be used towards the UV finiteness of
the perturbative supergravity \cite{Kallosh:2011dp}. In the $N=4$ case, where
there is a 1-loop global $U(1)$ duality anomaly \cite{mar}, one might
have some concerns regarding the  explanation of the 3-loop UV
finiteness in pure $N=4$ supergravity\cite{Kallosh:2012ei}. However, we
will argue below that in the underlying superconformal $N=4$ model the
local superconformal symmetry is anomaly free.

\section{Conformal compensator in $N=0$ supergravity}
Consider a model of pure gravity, $N=0$, promoted to a local Weyl conformal
symmetry:
\begin{equation}
S^{conf}= {1\over 2} \int d^4 x \sqrt {-g }\left( \partial_\mu \phi\,
\partial_\nu \phi \,g^{\mu\nu} + {1\over 6} \phi^2 \, R
 \right) .
\label{conf1}\end{equation} The field $\phi$ is referred to as a
conformal compensator. Various aspects of this  toy model of gravity with a Weyl compensator field (\ref{conf1}) were studied over the years \cite{Deser:1970hs}.
 The action is conformal invariant under the following local Weyl
transformations:
\begin{equation}
 g_{\mu\nu}'= e^{-2\sigma(x) } g_{\mu\nu}\,, \qquad
\phi'=e^{\sigma (x)}\phi  \,.\label{scaletransf}
\end{equation}
The gauge symmetry (\ref{scaletransf}) with one local gauge parameter can
be gauge fixed. We may  choose the unitary gauge
\be
\phi^2={6\over \kappa^2}
\ee
 Note
that one has to take a scalar field with ghostlike sign for the kinetic
term to obtain the right kinetic term for the graviton.  This does not
lead to any problems since this field disappears after the gauge fixing
and
 the action  (\ref{conf1}) reduces to the
Einstein action, which is not conformally invariant anymore:
\begin{equation}
S^{conf}_{gauge-fixed}= \frac{1}{2\kappa^2} \int d^4 x \sqrt {-g }\,
R
\,.
\end{equation}
 In
this action, the transformation (\ref{scaletransf}) does not leave the
Einstein action invariant any more. The $R$ term transforms with
derivatives of $\sigma(x)$, which in the action (\ref{conf1}) were
compensated by the kinetic term of the compensator field and  the weight was compensated by the $\phi^2$ term which is not present in the gauge-fixed action anymore. But the general covariance is still the remaining local symmetry of the action.

Now let us look for  the consequences of our conjecture that the  local
(super)conformal symmetry is fundamental, instead of Poincar{\'e} (super)gravity.

1.
We know that the first UV divergence that was predicted in pure gravity (not taking into account the conformal predictions, but only general covariance) at the 2-loop level   \cite{Kallosh:1974yh} is given by the cube of the Weyl tensor
\be
\Gamma_{2}^{N=0} \sim {1\over \epsilon} \kappa^2 \int d^4 x \sqrt {-g}  \, C_{\mu\nu}{}^{\lambda\delta} C_{\lambda\delta}{}^{\alpha \beta}   C_{\alpha\beta}{}^{\mu\nu} \ .
\label{prediction}
 \ee
Here the  on-shell condition is
$
R=R_{\mu\nu}=0$ and $ C_{\mu\nu\lambda \delta} = R_{\mu\nu\lambda \delta}
$.

2.  The actual computation was performed  in Ref. \cite{Goroff:1985th}, which demonstrated that 2-loop gravity is indeed UV divergent:
\be
\Gamma_2^{N=0}=  {\kappa^2\over (4\pi)^4 } {209\over  2880} \, {1\over \epsilon}\,  \int d^4 x \sqrt {-g}  \, C_{\mu\nu}{}^{\lambda\delta} C_{\lambda\delta}{}^{\alpha \beta}   C_{\alpha\beta}{}^{\mu\nu}  \ .
\label{prediction1}\ee

This finalized a convincing story of the UV infinities in pure $N=0$ quantum gravity. There is no reason to expect that the 3-loop
counterterm,
as well as all higher loop order  ${1\over \epsilon}$ UV divergences, will not show up.

Now assume that we use the underlying conformal model with local conformal symmetry. It is easy to promote the 2-loop UV divergence to the form of a conformal invariant:
\be
 \int d^4 x \sqrt {-g}  \, \phi^{-2} \, C_{\mu\nu}{}^{\lambda\delta} C_{\lambda\delta}{}^{\alpha \beta}   C_{\alpha\beta}{}^{\mu\nu}  \ .\ee
 Upon gauge-fixing it will produce the candidate for the 2-loop divergence.
Thus, even if we would use the embedding of gravity into a model with conformal symmetry by introducing an extra scalar compensator,  it would not help us to forbid the 2-loop UV divergence in the $N=0$ supergravity.

\subsection{$N=0$ supergravity with matter}

If we would add some additional matter to our  super\-conformal $N=0$ toy model, we would have to consider the 1-loop conformal counterterm independent on the compensator field, proportional to the square of the Weyl tensor
\be
\Gamma_{1}^{N=0} \sim {1\over \epsilon} \int d^4 x \sqrt {-g}  \, C_{\mu\nu\lambda\delta} C^{\lambda\delta\mu \nu}    \ .
\label{prediction1loopM}
 \ee
and in the topologically trivial background this counterterm is
\be
\Gamma_{1}^{N=0} \sim {2\over \epsilon} \int d^4 x \sqrt {-g}  \, \Big ( R_{\mu\nu} R^{\mu\nu} - {1\over 3} R^2  \Big )  \ .
\label{prediction1loop}
 \ee
The coefficient in front depends on the matter content. The reason for its absence in pure gravity was explained using the background field method in Ref. \cite{'tHooft:1974bx} by the fact that it is proportional to classical equations of motion
when the right-hand side (rhs) of the Einstein equation, $T_{\mu\nu}^{mat}$ is vanishing.
This means that the relevant divergence can be removed by change of variables.
In Ref. \cite{Kallosh:1978wt} it was shown explicitly that in pure gravity there is a choice
of the parameters $a, b$ of the general covariance gauge-fixing condition, of the type $a D_\mu h^{\mu\nu} -bD_\mu h=0$
which makes both the UV divergences $R_{\mu\nu} R^{\mu\nu}$ and $R^2$ $a,\, b$ gauge-dependent, and vanishing at certain values of $a,\,b$.

Thus, in the exceptional case of pure gravity without matter there is a  1-loop UV finiteness; the same is valid for all pure supergravities without matter. However, in presence of matter in gravity as well as in supergravities, the 1-loop UV divergences that are present are defined by the matter part of the energy momentum tensor.
\be
\Gamma_{1}^{N=0} = {1\over \epsilon} \int d^4 x \sqrt {-g}  \, \Big ( \alpha  (T_{\mu\nu}^{mat})^2+  \beta (T^{mat})^2  \Big )  \ .
\label{prediction1loop2}
 \ee
where $\alpha$ and $\beta$ depend on the matter content of the given model.

\section{$N=1$, $N=2$, $N=4$  supergravity}
\subsection{$N=1,2$}
The generic $N=1,2$ supergravity models were derived by gauge fixing the $N=1,2$ superconformal algebra, starting with $SU(2,2|1),  SU(2,2|2)$, respectively
(see Ref. \cite {TheBook} and references therein). The known facts are

1. In $N=1,2$ supergravities the prediction was made in Ref. \cite{Deser:1977nt} that the 3-loop divergence of the form
\be
\Gamma_{3}^{N=1,2} \sim {1\over \epsilon}\kappa^{4} \int d^4 x\,  \sqrt{-g} \,    C_{\alpha \beta \gamma \delta}   C_{\dot \alpha \dot \beta \dot \gamma \dot  \delta}  C^{\alpha \beta \gamma \delta}    C^{\dot \alpha \dot \beta \dot \gamma \dot  \delta}
\label{DS}\ee
is possible.

2. There were no computations of the 3-loop UV divergence in $N=1$, $N=2$ supergravity so far.

 The prediction in (\ref{DS}) was based on local supersymmetry, associated with Poincar{\'e} $N=1,2$ supergravity.

 The superconformal embedding prediction would require us to provide the superconformal embedding of the term in (\ref{DS}). The question is: is there a $N=1,2$ superconformal generalization of the expression
 \be
 \int d^4  x \,  \sqrt{-g} \,   \phi^{-4}  C_{\alpha \beta \gamma \delta}   C_{\dot \alpha \dot \beta \dot \gamma \dot  \delta}  C^{\alpha \beta \gamma \delta}   C^{\dot \alpha \dot \beta \dot \gamma \dot  \delta} \ ,
\label{DSconf}\ee
which is the  gravity part of the full $N=1,2$ superconformal higher derivative invariant? The answer is positive and is based on the fact that in $N=2$ there is a local superconformal calculus and there are  chiral multiplets with arbitrary Weyl weight \cite{Toine}, in particular
the negative powers of the compensator multiplet, which can be used for building higher derivative superconformal invariants. Moreover,
 various examples of superconformal higher derivative invariants in the $N=2$ model are presented in Ref. \cite{deWit:2010za}
 and recently used for comparison with on-shell superspace counterterms in Ref. \cite{Chemissany:2012pf}. The simplest $N=2$ superconformal version of  $R^4$  corresponding to minimal pure $N=2$ supergravity is given by the following chiral superspace integral
 \cite{Chemissany:2012pf}:
\be
 \lambda \int d^4\theta \left( {W^2\over S^2} \mathbb{T} \left({\overline{W^2} \over \overline{S^2}}\right)\right)\,.
  \label{faction}
\ee
 The $N=2$ superconformal calculus allows us to use the chiral multiplets $S^{-2}$ as well as any higher negative power $S^{-2n}$
 for building higher and higher derivative invariants in the $N=2$ supergravity   \cite{deWit:2010za}. Thus the hidden local superconformal $N=2$ symmetry does not lead to a particular restriction on $N=2$ supergravity counterterms.

\subsection {$N=4$ supergravity}

1.  The prediction was made in Ref. \cite{Bossard:2011tq}  that the 3-loop divergence of the form
\be
\Gamma_{3}^{N=4} \sim {1\over \epsilon}\kappa^{4} \int d^4 x\,  \sqrt{-g} \,    C_{\alpha \beta \gamma \delta}   C_{\dot \alpha \dot \beta \dot \gamma \dot  \delta}  C^{\alpha \beta \gamma \delta}    C^{\dot \alpha \dot \beta \dot \gamma \dot  \delta}
\label{DSN4}\ee
is expected since the relevant candidate counterterm has all required nonlinear symmetries of $N=4$ supergravity,
including the $SU(1,1)\times SO(6)$ duality.

2. The recent computation in Ref. \cite{Bern:2012cd} revealed that
\be
\Gamma_{3}^{ N=4} =0 \ .
\label{finite}\ee

The computations of UV loop divergences in Ref. \cite{Bern:2007hh} and in Ref. \cite{Bern:2012cd} are based on the information about the tree
amplitudes and on the unitarity method. Therefore these computations
seem to shed some light on all version of extended supergravity, which
at the tree level are equivalent. Such versions are related by various
classical duality transformations.
 We proceed from here by suggesting a conjecture of a hidden superconformal symmetry, which these computations may have revealed.

We will now proceed with the analysis based on our conjecture that the local superconformal supersymmetry may control the UV divergences of $N=4$ Poincar{\'e} supergravity.

 \section{ $N=4$ superconformal symmetry and supergravity}
Here we follow \cite{Bergshoeff:1980sw,Bergshoeff:1980is}
 and specifically \cite{deRoo:1984gd}, where the details on $N=4$ case have been worked out. To derive $N=4$ supergravity from the superconformal model
based on the $SU(2,2|4)$ graded algebra
 requires a number of rather complicated steps. We will only describe here the ones that are relevant for our purpose, referring the reader to the original papers
 \cite{Bergshoeff:1980sw,Bergshoeff:1980is,deRoo:1984gd}.

To derive the action of a pure $N=4$ Poincar{\'e} supergravity one has to start with 6 (wrong sign) metric $N=4$ vector multiplets
interacting with the $N=4$ Weyl gravitational multiplet. The Abelian vector multiplet action with the correct
sign of the metric invariant under rigid $N=4$ supersymmetry is
\be
-{1\over 4} F_{\mu\nu} F^{\mu\nu }  - \bar \psi^i \gamma \cdot \partial \psi_i -{1\over 2}\partial _\mu\phi_{ij} \partial^\mu \phi^{ij}
\ ,
\ee
with $ i,j=1,...,4$. For the six compensator vector multiplets ($I,J=1,\ldots ,6$) we take
\be
-{1\over 4} F_{\mu\nu}^{I} \eta_{IJ} F_{\mu\nu}^{J} - \bar \psi^{iI}\eta_{IJ} \gamma \partial \psi_i^J -{1\over 2}\partial _\mu\phi_{ij}^I \eta_{IJ} \partial_\mu \phi^{ij
J}\ ,
\ee
where $\phi ^{ij}=(\phi
  _{ij})^*=\varepsilon ^{ijk\ell }\phi _{k\ell }$ and
the constant real metric $\eta_{IJ}$ is diagonal and has six negative eigenvalues, $- 1$. This action is invariant under global $SU(4)$. The six negative eigenvalues point towards the role of $\phi_{ij}^I$ as compensators of a conformal symmetry, as explained in the toy model above. In the case of pure $N=4$ supergravity without matter multiplets all scalars from the six $N=4$ superconformal vector multiplets are the compensator scalars, as we will see below.

To derive the $N=4$ pure supergravity action one starts with six such vector multiplets and couple them to the fields of conformal $N=4$ supergravity.
There is a derivative ${\cal D}_a= e_a^\mu {\cal D}_\mu$, which is covariant with respect to all superconformal symmetries of $SU(2,2|4)$. Meanwhile ${\cal D}_\mu$  is covariant under Lorentz, Weyl, $SU(4)$ and $U(1)$ symmetries.
The $S$ and $K$ covariantization is performed in Ref. \cite{deRoo:1984gd} explicitly.

The  rigid supersymmetry algebra  $\{ Q, Q\}$ leads to translation $P$, so it is necessary to convert it into
general coordinate transformations to describe the coupling with gravity. This and analogous steps require  some constraints on the curvatures as well as introduction of fields, in addition to gauge fields above, to close the algebra, so that, after all
\bea
[ \delta_Q(\epsilon_1), \delta _Q(\epsilon_2)] &=&\delta_G^{\rm cov}(\xi^\mu)+ \delta_M (\epsilon^{ab})+\delta_Q(\epsilon_3^i)\nonumber\\
&& +\delta_S (\eta^i)
 + \delta_{SU(4)}(\lambda^i{}_j)+
\delta_{U(1)}(\lambda_T)\nonumber\\ && +\delta_K(\lambda_K^a) + \delta_A(\lambda)  +
X_{EOM }
\label{com}
\eea
The rhs of this commutator depends on a
combination of all local symmetries: general covariant, Lorentz,
supersymmetry, special supersymmetry, $SU(4)$, $U(1)$, conformal boosts, Abelian gauge
transformation on the vector fields, and spinor equation of motion on
$\psi^i$, which we show in the last term in
  $X_{EOM }$. The explicit expressions are derived in Ref. \cite{deRoo:1984gd}. The parameters of all these transformations, which form an `open algebra', are bilinear in $\epsilon_1(x)$, $\epsilon_2(x)$. The constraints on the curvatures lead to certain relations between the gauge fields so that some of them are not independent anymore.

\subsection{Superconformal coupling of vector multiplets to the Weyl multiplet}
The $N=4$ superconformal Lagrangian of the vector multiplets interacting with the Weyl multiplet is given in Eq. (3.16) in Ref. \cite{deRoo:1984gd} and takes a full page. We will present here the bosonic part of the action
for the six compensating vector multiplets, with the wrong sign of kinetic terms, which is relatively
simple:
\bea
e^{-1} L^{bos}_{s.c.}&=& -{1\over 4} F_{\mu\nu}^{+I} \eta_{IJ} F_{\mu\nu}^{+J}{\phi^1-\phi^2\over \Phi}   -{1\over 4}{\cal D} _\mu\phi_{ij}^I \eta_{IJ}{\cal D} _\mu \phi^{ij J} \nonumber \\
\nonumber \\
&-&F_{\mu\nu} ^{+I }\eta_{IJ}T^{\mu\nu}_{ ij}\phi^{ijJ} {1\over \Phi}
-{1\over 2} T_{\mu\nu ij}\phi^{ijJ} \eta_{IJ} T^{\mu\nu}_{ kl}\phi^{klJ}{\Phi^*\over \Phi}\nonumber \\
\nonumber \\
&-&{1\over 48} \phi_{ij}^I \eta_{IJ} \phi^{ij J} \Big ( E^{kl}E_{kl}+ 4 D_a\phi^\alpha D^a\phi_\alpha  - 12 f_\mu{}^\mu   \Big) \nonumber \\
\nonumber \\
&+&{1\over 8} \phi_{ij}^I \eta_{IJ} \phi^{kl J} D^{ij}{}_{kl}
+h. c.
\label{action}
\eea
Here
\be
 \Phi=\phi^1+\phi^2\ , \qquad \Phi^*=\phi_1-\phi_2 \ , \qquad \phi^\alpha \phi_\alpha =1 \ .
 \label{defPhiphi}
\ee
The conformal boost gauge field $f_\mu{}^a$ is a function of a curvature
\be
f_\mu{}^\mu= - {1\over 6 } R(\omega)\ .
\ee
The scalars $\phi_\alpha$ with $\alpha=1,2$ transform as a doublet under $SU(1,1)$.
The constraint (\ref{defPhiphi}) and the $U(1)$ gauge invariance   reduce the 2 complex variables $\phi _\alpha $ to 2 real fields, so that they are in $SU(1,1)\over U(1)$ coset space.

The fields $T_{\mu\nu ij}, E^{kl}, D^{ij}{}_{kl}$ belong to the gravitational Weyl multiplet. This action, supplemented by all fermionic terms, leads to
equations of motion of the fermion partner of the compensator scalars $\psi^{iI}$ in the rhs of the commutator of two supersymmetries (\ref{com}), which we denoted by $X_{EOM }$.

 The action is linear in $D^{ij}{}_{kl}$, so it is convenient for the purpose of future gauge fixing to use the reparametrization of the 36
  variables $\phi _{ij}{}^I$ in terms of the  36 $ \varphi_M{}^I(x)\equiv\{ \varphi_m^I(x), \varphi _{m+3}^I (x)\}$, so that
$M=1,...,6$:
 \be
 \phi_{ij}^I (x)= \varphi_m^I(x) \beta^m{}_{ij} + {\rm i} \varphi _{m+3}^I (x) \alpha^m{}_{ij} \ ,
 \label{Sergio}
 \ee
where $\alpha^m$ and $\beta^m$ with $m=1,2,3$ are $SU(2)\times SU(2)$ numerical matrices introduced in Ref. \cite{Cremmer:1977tt}.

\section{Poincar{\'e} gauge}
The superconformal action has unbroken local $K$ , $D$, and $S$ symmetries which are not present in supergravity and must be
gauge fixed to convert the superconformal action into a supergravity one.\footnote{This is a precise analog of three gauge symmetries in the $SU(3)\times SU(2)\times U(1)$ model which have been gauge fixed in the unitary gauge where $W^{\pm}$ and $Z$ are massive vector mesons.} This is done the same way as in the toy example above, namely, the scalar compensator dependent term in front of $R$ is designed to introduce a Planck mass into conformal theory which
originally, before gauge fixing, has no dimensionful  parameters. To fix the local dilatation $D$ one can take
\be
 \phi_{ij}^I \eta_{IJ} \phi^{ij J}=-{6\over \kappa^2} \ .
\ee
This provides the Einstein curvature term in the action
\be
-{1\over 24}  \phi_{ij}^I \eta_{IJ} \phi^{ij J} R \quad \Rightarrow \quad  {1\over 4 \kappa^2} R
\ee
and explains why the diagonal metric $\eta_{IJ}$ has six negative values.
The  $S$ and $K$ local symmetries are fixed by taking
\be
b_\mu=0 \ , \qquad  \psi_{i}{}^{ J}=0 \ .
\label{gaugedilatonK}
\ee
The fact that our six vector multiplets have a wrong sign kinetic terms is in agreement with the fact that the scalars are conformal compensators.
As long as $\varphi_m^I(x)$ and $\varphi _{m+3}^I (x)$ with $m=1,2,3$ are present, there is also a local $SU(4)$ symmetry.
So, we can use the 15 parameters from $SU(4)$ together with $20+1$ conditions --- field equations of $D^{ij}{}_{k\ell }$ and the
  dilatation gauge mentioned in (\ref{gaugedilatonK}) --- to take
 the scalars in Eq. (\ref{Sergio}) constant,
 \be
\varphi_M{}^I(x)= {1\over 2\kappa} \delta_M{}^I \ ,
\ee
to remove these 36 variables.

The remaining important steps include the elimination of the auxiliary fields of the Weyl multiplet, $E^{kl}$ and $T_{\mu\nu ij}$.
The field $E^{kl}$ turned out to be  proportional to fermion bilinears, which changes the fermionic part of the action.
However, the role of $T_{\mu\nu ij}$ is extremely important: the procedure of  its exclusion on its equations of motion
leads to a sign conversion of the kinetic term for the vectors from the six vector multiplets---they become physical vectors
with the correct sign kinetic term.

To summarize, the six vector multiplets at the superconformal stage all have wrong kinetic terms since the $N=4$ scalar partners play
the role of conformal compensators. When scalars are gauge fixed to eliminate the local Weyl $D$ symmetry (dilatation), the Einstein gravity arises.
The six quartets of spinors ($6\times 4\times 4=96$ components) from the vector gauge multiplets are eliminated by the combination
of 16 gauge conditions of  local $S$ supersymmetry
(special supersymmetry) and  the field equations of the auxiliary fermions in the Weyl
  multiplet (80 components).

The vectors from the six vector
multiplets are converted into physical vectors of supergravity, when the auxiliary field $T_{\mu\nu ij}$ of the Weyl multiplet
is excluded on its equations of motion. The action becomes that of pure $N=4$ supergravity in $\kappa^2=1$
units where the local $U(1)$ symmetry is still present. The bosonic part is
\bea
&&e^{-1} L^{\rm bos}_{\rm sg}= {1\over 4 } R(\omega) + {1\over 2 } D_a\phi^\alpha D^a\phi_\alpha +
\nonumber\\
 %\nonumber\\
&&+{1\over 4} F_{\mu\nu}^{+I} \eta_{IJ} F^{+J \mu\nu}\, {\phi_1+\phi_2\over \phi_1-\phi_2}+h. c.
\label{sugra}
\eea
Note that the actions are supersymmetric after adding
the fermionic part. This means that the variation of the action vanishes
for arbitrary field
configurations: the fields do not satisfy any equations. The statement
that the multiplets are ``on shell'' is a statement on the algebra of
transformations, and that one depends on specific field equations. Thus
no other invariant can be constructed with these on shell
multiplets, since this would change the field equations. This is why the inverse powers of the vector multiplet (or a
logarithmic function)  cannot be used to construct other invariant
actions, see a further discussion of this in Sec. \ref{ss:higherderscN4}.

%\
%
%\noindent {\it Triangular  $U(1)$ gauge-fixing}
%
%\
\subsection{Triangular  $U(1)$ gauge-fixing}

\noindent The  local $U(1)$ gauge
 we take\footnote{The related construction is discussed in Ref. \cite{deRoo:1985jh}, without making an explicit choice of the $U(1)$ gauge.}
 is
\be
{\rm Im} \, (\phi_1-\phi_2) =0
\ee
Our choice is motivated by the triangular decomposition of the $SL(2,\mathbb{R})$ matrix of our model. We start with the $SU(1,1)$ matrix defined in Ref. \cite{deRoo:1984gd}
\be
\label{U} U
 = \left(
                      \begin{array}{cc}
                     \phi_1&  \phi_2^* \\
                     \phi_2 & \phi_1^* \\
                      \end{array}
                    \right)
\ee
We  switch to the $SL(2,\mathbb{R})$ basis
and get
\be
\label{sl2r} S
 ={\cal A}\, U \,  {\cal A}^{-1}=  \left(
                      \begin{array}{cc}
                     \Re (\phi_1+\phi_2)& -  \Im (\phi_1+\phi_2) \\
                     \Im (\phi_1-\phi_2) & \Re (\phi_1-\phi_2) \\
                      \end{array}
                    \right)
\ee
where
\be
\label{A}
{\cal A} = {1\over \sqrt 2} \left(
                      \begin{array}{cc}
                      1&  1 \\
                     -i & i \\
                      \end{array}
                    \right)
\ee
We define an  independent variable $\tau$  as
\be
 \tau = \tau_1+i \tau_2 \equiv i {\phi_1+\phi_2\over \phi_1-\phi_2}
\ee
which parametrizes the coset space ${SL(2,\mathbb{R})\over U(1)}$.
We take
\be
\phi_1= {1\over 2 \sqrt{\tau_2}}(1-i\tau) \qquad \phi_2=- {1\over 2 \sqrt{\tau_2}}(1+i\tau)
\ee
In this notation with $\tau_2= e^{-2\varphi}$ and $\tau_1=\chi$ the triangular decomposition of the $SL(2,\mathbb{R})$ matrix  is clear
\be
\label{sl2rtring} S
 =  \left(
                      \begin{array}{cc}
                    e^{-\varphi}& \chi e^{\varphi} \\
                    0 & e^{+\varphi} \\
                      \end{array}
                    \right)
\ee
When these values of $\phi_\alpha$ are inserted into the superconformal action (3.16)  or the partially gauge-fixed (4.18)
of \cite{deRoo:1984gd},  we   get the Cremmer-Scherk-Ferrara (CSF) $N=4$ supergravity model \cite{Cremmer:1977tt} the bosonic part of which is
%\begin{eqnarray}
%&& e^{-1} L_{CSF}
%= {1\over 2 } R -{1\over 4} {\partial \tau \partial\bar \tau \over  (\Im \tau )^2} \nonumber\\
%&& +{1\over 4}\delta_{IJ} \left[i \tau  F_{\mu\nu}^{+I}  F^{+J \mu\nu}  +h. c.\right]
% \label{Lvsv}
%\end{eqnarray}
%
\begin{equation}
  e^{-1} L_{CSF}
= {1\over 2 } R -{1\over 4} {\partial \tau \partial\bar \tau \over  (\Im \tau )^2} +{1\over 4}\delta_{IJ} \left[i \tau  F_{\mu\nu}^{+I}  F^{+J \mu\nu}  +h. c.\right]
 \label{Lvsv}
\end{equation}

%\
%
%\noindent {\it Bergshoeff, de Roo, de Wit  $U(1)$ gauge}
%
%\

\subsection{Bergshoeff, de Roo, de Wit  $U(1)$ gauge}

\noindent The choice in Ref. \cite{Bergshoeff:1980is,deRoo:1984gd} is
\be
\Im \, \phi_1= 0
\ee
and the independent variable is defined as
\be
Z\equiv  {\phi_2\over \phi_1}\ee
Here the scalars
\be
\phi_1= {1\over \sqrt{1-|Z|^2}}\ , \qquad \phi_2= {Z\over \sqrt{1-|Z|^2}}  \ ,
\ee
parametrize the coset space ${SU(1,1)\over U(1)}$.
The six vector multiplets have kinetic terms with the correct sign. The theory has a global duality symmetry $SU(1,1)\times SO(6)$ inherited from the superconformal $N=4$ model. The scalar  couplings are
\be
- {\partial Z \partial \bar Z\over( 1-|Z|^2)^2}
\ee
In this local $U(1)$ gauge the $N=4$ supergravity is an intermediate version between the  CSF model \cite{Cremmer:1977tt} and the one given in Ref. \cite{Das:1977uy} and in full details in Ref. \cite{Cremmer:1979up}. Specifically, the scalars $Z$ are the same as in Ref. \cite{Cremmer:1979up}, however, the vectors are related to the ones in Ref. \cite{Cremmer:1979up} by a duality transformation.

Thus we find that our new gauge which provides a CSF $N=4$ supergravity model \cite{Cremmer:1977tt} directly from the superconformal model is nice and simple, comparative to other versions of $N=4$ supergravity.

\section{Higher derivative superconformal actions in $N=4$ model}
\label{ss:higherderscN4}

There is only one type of possible matter multiplets in $N=4$ supersymmetry, i.e. $N=4$ Maxwell multiplets (and the non-abelian version).
The scalar $\phi_{ij}$ has Weyl weight $w=1$ and therefore it is used to gauge fix local dilatation by the $\phi_{ij}^I \eta_{IJ} \phi^{ij J}=-{6\over \kappa^2} $ condition.
The fact that the algebra does not close on these $N=4$ Maxwell  multiplets implies that we cannot use them anymore in further tensor calculus for $N=4$. Moreover, the local superconformal symmetry algebra is closed on the Weyl multiplet.
Since there are no other multiplets in that case, it is not possible to construct the $N=4$ superconformal version of $C^4$ shown in Eq. (\ref{DSconf}), which requires a superfield with the conformal weight $w=-4$ which in the Poincar{\'e} gauge becomes $\kappa^4$.
It would require an $N=4$ superconformal version of the bosonic expression $( \phi_{ij}^I \eta_{IJ} \phi^{ij J})^{-2}C^4$,  which does not exist. Including supercovariant derivatives $D_a= e_a^{\mu} D_\mu$ with $w (e_a^{\mu})=+1$
can only increase the positive conformal weight $w$ of the corresponding superconformal invariant, which requires higher negative powers of a compensator.

The situation in the $N=4$ case is in sharp contrast with $N=1,2$ cases  where there are chiral superfields of  arbitrary conformal weight $w$
\cite{Toine},
which can be used to build the superconformal invariants. Moreover,  according to Eq. (C.2) in Ref. \cite{deWit:2010za} one can take an arbitrary  function of the chiral compensator superfield ${\cal G}(\phi)$ and construct such a negative conformal weight superfield out of the compensators
in the $N=2$ superconformal case.
In the Poincar{\'e} gauge such a superfield ${\cal G}(\phi)= \phi^{-2n}$ will provide the increasing powers of gravitational coupling $\phi^{-2n} \Rightarrow \kappa^{2n}$.

To explain why in $N=4$ superconformal theory it is not possible to produce superinvariant actions with arbitrary function of superfields consider an example : the off-shell chiral multiplet \be
(z,\chi_L, F)
\ee
We want to construct an action
$S = \int d^2\theta   {\cal G}(z)$. To find the components, we obtain the
fermion component of ${\cal G}$ by calculating one supersymmetry (SUSY) transformation on the lowest component ${\cal G}(z)$.
This gives ${\cal G}'(z) \chi_L$. A further transformation gives (for the component that will be integrated)
\be
{\cal G}'(z) F - (1/2) {\cal G}''(z) \bar\chi_L\chi_L
\ee
This transforms to $\gamma^\mu\partial_\mu [ {\cal G}'(z) \chi_L ]$
and thus gives a good invariant action.

However, consider now that we would have only the on-shell multiplet, e.g. for a massless multiplet. Then $F=0$.
The algebra on $\chi_L$ leads then to the field equation $\gamma^\mu\partial_\mu \chi_L  =0$.
The second SUSY transformation as above leads to
$- (1/2) {\cal G}''(z) \bar\chi_L\chi_L$
Thus the superfield would be
 \be
 {\cal G}(z) + \bar \theta_L {\cal G}'(z) \chi_L - (1/2) {\cal G}''(z) \bar \theta_L \theta_L \bar\chi_L\chi_L
 \ee
This is a superfield for any $ {\cal G}(z)$. However, the integral $\int d^2 \theta$  gives the last component,
which transforms under SUSY to
\be
\gamma^\mu \chi_L \partial_\mu  {\cal G}'(z)
\label{onshellToine}\ee
This is not a total derivative (missing a term proportional to the field equation, but that we cannot use to have an invariant action).
This illustrates that the multiplet calculus can only be used for off-shell multiplets.  We provide a more detailed discussion and relation to $N=2$ deformation in models with higher derivatives in the Appendix.

Before we take seriously a prediction on higher derivative superinvariants following from the local $N=4$ superconformal theory,
we have to study the situation with anomalies. The local anomalies for $N=1$ superconformal theories were studied in Ref. \cite{deWit:1985bn}
and in Refs. \cite{Schwimmer:2010za,Buchbinder:1988yu}.
Our $N=4$ superconformal model of six (wrong sign) compensators interacting with the Weyl multiplet, upon gauge-fixing,
leads to pure $N=4$ supergravity with Einstein curvature, without the square of the curvature in the action.
The local anomalies of this model will be discussed below.

 \subsection{Superconformal anomalies}

 Local superconformal anomalies were studied in detail in the $N=1$ case in Ref. \cite{deWit:1985bn}. It was explained there that the consistent anomalies can be constructed using the Wess-Zumino method \cite{Wess:1971yu}. In gauge theories the method allows us to construct terms $\Gamma(\Phi, A_\mu)$, whose variation takes a form of a consistent anomaly $\delta_\Lambda \Gamma(\Phi, A_\mu)$, which does not depend on the compensator field $\Phi$.
 Later the related work was performed in  a somewhat different context in Ref. \cite{Schwimmer:2010za} based on Ref. \cite{Buchbinder:1988yu}.
We are interested in local symmetry anomalies, which in gauge theory
examples may be fatal and lead to a quantum  inconsistent theory. For
example, the triangle local chiral symmetry anomaly in standard model,
if not compensated,  means that the physical observables  in the unitary
gauge do not coincide with the physical observables  in the
renormalizable gauge. The change of variables in the path integral of
the kind performed in Ref. \cite{Kallosh:1972ap}, which allows us to prove an
equivalence theorem for the S matrix in arbitrary gauges, may be
invalidated in presence of anomalies.

For example, in  the simple case of the $SU(2)$ gauge model we may be
interested in transverse renormalizable gauge $\partial^\mu A_\mu{}^{
m}=0$ or in the unitary gauge $ B^m=0$. To find the relation between
these two gauges one may look at a more general class of gauges like  $a
\, \partial^\mu A_\mu{}^{ m}+ b \, B^m=0$. In the unitary gauge the
theory is  not renormalizable off shell, however, if the equivalence
theorem
\be
\langle| S | \rangle |_{a,b}= \langle | S | \rangle |_{a+\delta a,b+\delta b}
\label{Tyutin}
\ee
is valid, the physical observable are
the same as the ones in renormalizable gauge (with account of some
dependence on gauge-fixing of renormalization procedure). Also the proof
of unitarity in the renormalizable gauge is based on the validity of
(\ref{Tyutin}).

The local symmetry anomaly may invalidate the $a, b$ independence  of physical observables. Instead of equivalence we have a relation
\begin{eqnarray}
 \langle| S | \rangle |_{a,b}  & = & \langle | S | \rangle |_{a+\delta a,b+\delta b} \nonumber\\
   &   & + X\langle \int   \Lambda ^\alpha(x, \phi^i, \delta a,\delta b) \,  {\cal A}_\alpha (\phi^i)\rangle\,.
 \label{X}
\end{eqnarray}
Here $ {\cal A}_\alpha (\phi^i)$ is the consistent anomaly depending on
various fields $\phi^i$ of the model, and $\Lambda ^\alpha(x, \phi^i,
\delta a,\delta b)$ is a specific change of variables, leaving
the classical action invariant, but effectively changing the gauge-fixing condition,
with examples given in Ref. \cite{Kallosh:1972ap}. $X$ is a numerical value
in front of a candidate anomaly, which may vanish, in the case of
cancellation,  or not, depending on the model.

 Thus, in the context of local anomalies which may exist and destroy the quantum consistency of the model, we will look at possible candidates for anomalies given by expressions like
 \be
 \delta_\Lambda  \Gamma (\phi^i)=\int d^4x\,  \Lambda ^\alpha(x) \,  {\cal A}_\alpha
 (\phi^i)\,,
\label{anomaly} \ee
where $ \Lambda ^\alpha(x)$ corresponds to all gauge symmetries of a given model.

 There are two conditions for an anomaly to be fatal for a gauge theory, i. e., to make quantum theory inconsistent.

 I. The candidate consistent anomaly (\ref{anomaly}) should be available according to local symmetries of the model

II. The numerical coefficient in front of a candidate anomaly, which is due to contribution from  various fields of the model, should not cancel, $X\neq 0$ in (\ref{X}).

\

{\it $N=1$ case}

The symmetries of $N=1$ superconformal models include
\be
\epsilon (x)\, ,  \qquad \eta(x)\, , \qquad \lambda_D(x)\, ,  \qquad
\lambda_T(x),
\label{param}\ee
i.e., local $Q$ supersymmetry, local $S$ supersymmetry, Weyl local conformal symmetry, local chiral $U(1)$ symmetry, respectively, and of course, general covariance and Lorentz symmetry.

In case of $N=1$ superconformal models the corresponding $\delta_\Lambda  \Gamma (\phi, W^2)$ was given in Ref. \cite{deWit:1985bn} in Eq. (5.7). The integrated   form of the anomaly is given by a local action in Eq. (5.6) in Ref. \cite{deWit:1985bn}
\be
  \Gamma^{dWG}  (\phi, W^2)\,.
\label{GdW}  \ee
Here $\phi$ is the compensator superfield of a Weyl weight $w=1$, and $W_{\alpha\beta\gamma}$ is a Weyl superfield of conformal weight $w=3/2$. The variation of   (\ref{GdW}) produces a consistent anomaly.  At the linear level this action is associated with the $F$ component of the chiral superfield
\be
  \Gamma^{dWG}  (\phi, W^2)  = (\ln \phi \, W_{\alpha\beta\gamma} W^{\alpha\beta\gamma})_F +...
\label{log}
\ee
Terms with ... involve important  corrections, required for locally superconformal action.
The superfield $\ln \phi \, W_{\alpha\beta\gamma} W^{\alpha\beta\gamma}$  seems to have a Weyl weight $w=3$, except that the  $\ln \phi$ does not have a uniform scaling weight $w=0$, which leads to complication  and modification of the scale, chiral and $S$-supersymmetry transformations. Nevertheless, the consistent exact nonlinear expression for $N=1$ superconformal anomaly in the form (\ref{anomaly})
\be
\delta   \Gamma^{dWG}  (\phi, W^2)
\label{gdwanoma}\ee
was established in Ref. \cite{deWit:1985bn} and given in Eq. (5.7) there.  It has terms with all local parameters in (\ref{param});
i. e., there is a Weyl local conformal symmetry anomaly, local chiral $U(1)$-symmetry anomaly, local $S$-supersymmetry anomaly and local $Q$-supersymmetry anomaly, all proportional to each other: either all of them or none. Note that the analysis in Ref. \cite{deWit:1985bn} was not based on specific computations of
anomalies;
it was an analysis based on consistency of the anomalies in $N=1$ superconformal models. The candidate consistent anomaly (\ref{anomaly})
is available; the coefficient $X$ in (\ref{X}) is model dependent.

It may be useful also to bring up here the relevant discussion of the
 $N=1$ superconformal anomaly in Refs. \cite{Schwimmer:2010za,Buchbinder:1988yu}.
The corresponding gauge-independent\footnote{We omit terms in Ref. \cite{Schwimmer:2010za} proportional to $R$ and $G_{\alpha \dot \beta}$
superfields as they depend on the gauge-fixing condition and may be
removed, as shown for example in Ref. \cite{Kallosh:1978wt}.} part of the
anomaly is given by
\be
  \Gamma^{ST} (\phi, W^2)= 2(c-a) \int d^8 z {E^{-1}\over R}  \,\ln  \phi  \, W_{\alpha\beta\gamma}
  W^{\alpha\beta\gamma}+c.c.
\label{Wloc}
\ee
and
\be
\delta   \Gamma^{ST} (\phi, W^2)= 2(c-a) \int d^8 z {E^{-1}\over R}  \,\delta \Sigma  \, W_{\alpha\beta\gamma}
W^{\alpha\beta\gamma}+c.c.,
\label{anom}
\ee
where under the superconformal transformations the compensator superfield transforms as
\be
\phi (x, \theta)  \rightarrow e^{ \, \Sigma (x, \theta)} \, \phi (x,
\theta)\,.
\ee
  Therefore, when its vacuum expectation value is non-vanishing, one may try to define the Goldstone superfield,
  which according to  \cite{Schwimmer:2010za} is ``dimensionless'' and transforms by a superfield shift
\be
\delta \ln \phi (x, \theta)  \rightarrow  \delta \Sigma (x, \theta)
\label{shift}\ee
and therefore
\be
\delta \ln \phi (x, \theta) W^2(x, \theta)=  \delta \Sigma (x, \theta)
W^2(x,\theta)\,.
\ee
Therefore, the local dilatations of the supermultiplet $\ln \phi$ are different
from those of a multiplet with a particular Weyl weight.
For example, for the chiral multiplet ($\phi =\{Z, \chi, F\}$) of the Weyl weight $w$ the superconformal transformations,
given for example in Eq. (16.33) in Ref. \cite{TheBook} have some $w$-dependent terms like
\be
\delta Z = w (\lambda_D + i \lambda_T) Z+...\\
\ee
where $\lambda_D(x) $ is a local dilatation and $\lambda_T(x) $ is a local chiral transformation. The same for $\chi, F$---there are terms depending on $w$.
These $w$-dependent terms are replaced by different transformations when
the fields do not scale homogeneously under local dilatations. These
transformations for $\ln \phi $ can be inferred from (\ref{shift}) where
\begin{equation}
  \Sigma =\left\{\lambda _D+i\lambda _T, \sqrt{2}\eta ,0\right\}
 \label{SigmaSConf}
\end{equation}
and corresponding changes in the superconformal derivatives.
The standard superconformal action for the multiplet, given for example
in Eq. (16.33) in Ref. \cite{TheBook}) is not superconformal invariant
anymore due to these corrections to the superconformal transformations.
However, its superconformal variation does not depend on the
compensator. As a result, the complete nonlinear expression for anomaly
in Eq. (5.7) in Ref. \cite{deWit:1985bn} is different from $\delta
\Gamma^{ST} (\phi, W^2)$ in (\ref{anom}). The expression for the anomaly
in (\ref{anom}) depends on manifestly $Q$-supersymmetric superfields and
gives the impression that only Weyl, chiral and $S$-supersymmetry
anomalies are consistent.  Meanwhile, the extra terms in $  \delta
\Gamma^{dWG}  (\phi, W^2)$  involve also the $Q$-supersymmetry anomaly,
and therefore the complete nonlinear expression for $N=1$ superconformal
anomaly is not given in terms of superfields with manifest
$Q$ supersymmetry, but in Eq. (5.7) in Ref. \cite{deWit:1985bn}.

Thus, a complete expression  in Eq. (5.7) in Ref. \cite{deWit:1985bn} for the
superconformal anomaly $\delta \Gamma^{dWG}  (\phi, W^2) $ of $N=1$
superconformal models contains  local  scale, chiral, $S$-supersymmetry
and $Q$-supersymmetry anomalies.  It is generated by the superconformal
variation of the expression $\Gamma^{dWG}  (\phi, W^2) $. This is a
construction of a consistent  anomaly which we intend to generalize to
the $N=4$ case.

 {\it $N=4$ case}

For the $N=4$ superconformal anomaly the  actions of the type
(\ref{log}) and
(\ref{Wloc}) are not available. The reason is the same as we have
already explained with regard to candidate counterterms. In the $N=4$ case the
generalization of the $N=1$ case of  arbitrary functions of a chiral
compensator like ${\cal G}(\phi)= \phi^{-2n}$ is not available; such superfields cannot be used to provide invariant actions. Indeed,
the compensating multiplets are in this case the vector multiplets whose
transformations close only using specific field equations. Therefore,
one cannot manipulate with these multiplets, as we explained in the beginning of this Sec. \ref{ss:higherderscN4}. This excludes also a
possibility to use ${\cal G}(\phi)=\ln \phi$ for building superinvariants.
Therefore there is no supersymmetric version of $\ln \phi (R-R^*)^2$ (for
chiral anomaly). It is available in $N=1$ and $N=2$ superconformal theories
but not available in $N=4$. The $N=4$ superconformally invariant  version of
$\ln ( \phi_{ij}^I \eta_{IJ} \phi^{ij J}) (R-R^*)^2$ is not available.

The restrictions of $N=4$ superconformal symmetry are significantly stronger than the ones for $N=1,2$.
$Q$ supersymmetry has a limited restriction on $N$-extended supergravity counterterms, and suggests that
for $N$-extended supergravity the $L=N$ geometric on-shell counterterms are
available;
the same prediction follows from $N=1,2$ superconformal models. However, for $N=4$,
the symmetries allow only the local classical action and protect the model from anomalies and counterterms.

This supports our conjecture that $N=4$ superconformal models are
quantum mechanically consistent and therefore we may trust the analysis
of candidate counterterms based on $N=4$ superconformal symmetry, which
predicts the UV finiteness of perturbative theory.

\subsection{Can we falsify our arguments using more general $N=4$ models?}
1. Consider $N=4$ supergravity interacting with some number $n$ of $N=4$
vector multiplets. The superconformal un-gauge-fixed version of this
model is described in Refs. \cite{deRoo:1984gd,deRoo:1985jh}. It
corresponds to the model which we present in Eq. (\ref {action}) where
$\eta_{IJ}$ has six negative eigenvalues as well as $n$ positive
eigenvalues.

There is a 1-loop UV divergence in the case of $N=4$ SG + $N=4$ vector
multiplets (see for example \cite{Fischler:1979yk}). There is also a
corresponding counterterm in the underlying superconformal theory; it
contains the square of the Weyl tensor.  The linearized version of it
 \footnote{The complete nonlinear bosonic action was recently derived in
Ref. \cite{Arkady} by  integrating over the  $N=4$  Yang-Mills fields.}
is given in Eq. (3.17) of \cite{Bergshoeff:1980is}. The complete
nonlinear action for the $N=2$ superconformal case is given in Eq.
(5.18) of \cite{Bergshoeff:1980is}.

The existence of this 1-loop counterterm is in agreement with $N=4$
supergravity analysis, as well as actual computations in Ref.
\cite{Fischler:1979yk}. In components is starts with
$C_{\mu\nu\lambda\delta}^2 +...$ which corresponds to   $R^2$ and
$R_{\mu\nu}^2$ terms, as explained in Eqs. (\ref{prediction1loopM}) and
(\ref{prediction1loop}). These vanish for pure $N=4$ supergravity,
corresponding to the model with six $N=4$ compensators,  since only in
pure supergravity $R=R_{\mu\nu}=0$. In the presence of matter
multiplets, the counterterm  has terms which do not vanish on shell,
like $T_{\mu\nu}^2$ and $T^2$.

The 1-loop $N=4$ square of the Weyl multiplet counterterm  is
superconformal by itself; it does not need $N=4$ compensators since it
has a correct Weyl weight. This is why it escapes the problem with a
negative power of compensators, which is present for all $N=4$
superconformal invariants, starting with 3 loops. They need
$\phi^{-2(L-1)}$ corresponding to $\kappa^{2(L-1)}$. Clearly, for L=1
there is no such dependence on a compensator.

 2. Now we apply our method to  $N=4$ conformal  supergravity interacting with some  $N=4$ vector multiplets \cite{Fradkin:1985am}.  This model is believed to be renormalizable but has ghosts.
The superconformal counterterm  corresponding to the square of the Weyl
tensor is not excluded and it is not vanishing. It  is not proportional
to the equations of motion of conformal supergravity interacting with
any number of vector multiplets. Thus the renormalizable UV divergences,
proportional to the part of conformal supergravity classical action are
expected.  And since again this particular unique superinvariant has the
proper Weyl weight, the action does not depend on compensators.
Therefore this counterterm escapes the problem with negative power of
compensators.

\subsection{Half-maximal D=6 superconformal models}

Maybe a  a simple counterargument to our conjecture
 comes from higher dimensions where known divergences occur
already in supergravity theories at low loop orders. For example,
maximal supergravity diverges in $D=6$ already at the 3-loop order, with
a $d^6R^4$ counterterm. It is possible that such 3-loop divergence takes
place also in the 16-supercharge half-maximal theory. If the classical
theory in $D=6$ can  be promoted to a compensated conformal supergravity
theory,  one would conclude that in an analogous situation our
conjecture is already proven to be invalid.

Here we analyze  supersymmetry and supergravity with 16 real
supercharges in $D=6$. From the properties of spinors, it follows that
we should divide this in $(2,0)$ and $(1,1)$ theories.

The $R$-symmetry group follows from the analysis of Jacobi identities as
in Sec. 12.2 of \cite{TheBook}. That leads to $\USp(4)=\SO(5)$ for the
$(2,0)$ and $\USp(2)\times \USp(2)$ for the $(1,1)$ theory.

Reductions of theories with 16 supercharges in higher dimensions on tori
cannot lead to chiral theories, and thus these lead to $(1,1)$. On the
other side, both theories lead to the same theories in five or fewer
dimensions, e.g. to  $N=4$ in $D=4$.

The $(1,1)$ theory allows vector multiplets (the reduction of the vector
multiplet of $D=10$). The $(2,0)$ theory does not have vector
multiplets, but has self-dual tensor multiplets. The quadratic action of
a two-form is conformal in six dimensions, and the self-dual tensor
multiplet can be defined with superconformal symmetry. However, there is
no action, only field equations, due to the self-dual
properties.\footnote{There are two ways around this obstruction: the
Pasti-Sorokin-Tonin method with  extra gauge symmetries and fields
\cite{Pasti:1996vs} and a ``pseudo-action'' \cite{Bergshoeff:2001pv},
from which field equations can be obtained after imposing self-duality
conditions. But in both cases the construction of the $(2,0)$ D=6
superconformal action has not been achieved.} The multiplet is an
on-shell multiplet. It is the quadratic approximation to the $M$5 brane,
and the superconformal structure of this multiplet was studied in Ref.
\cite{Claus:1997cq}.

The (1,1) supergravity has been constructed in Refs.
\cite{Giani:1984dw,Romans:1985tw}. This could not be done using
superconformal methods for the arguments that  will be now explained.

To start with a superconformal calculus we first need a superconformal
group. That group should contain the conformal group $\SO(6,2)$ and the
$R$-symmetry group as bosonic subgroup. In fact, we even need the
covering group of the conformal group since we need fermions. That
covering group is $\SO^*(8)$. All superconformal constructions have been
based on a simple superalgebra. We thus consider the list of real forms
of simple superalgebras, \cite{realLieSA} which can be also inferred
from Ref. \cite{VanProeyen:1999ni}. Then we consider which are the
possibilities between those that have the appropriate bosonic
subalgebra. This is what Nahm did in Ref. \cite{nahm}.

Then we see that with $\SO^*(8)$ only $\OSp(8^*|q)$ is available, which
identifies $\USp(q)$ as $R$-symmetry group. Thus there is no possibility
to have $\USp(2)\times \USp(2)$, as we would need for type (1,1)
supergravity, but there is $\USp(4)$, which is the $R$-symmetry group of
(2,0). Therefore only (2,0) allows a superconformal construction, based
on the superalgebra $\OSp(8^*|4)$. This was constructed\footnote{Note
that the theory was also constructed without superconformal methods in
Ref. \cite{Riccioni:1997np}.} in Ref. \cite{Bergshoeff:1999db}. In that
paper, $N+5$ tensor multiplets were coupled to the Weyl multiplet. Five
of these are compensating, leaving tensor multiplets with scalars in the
coset $\frac{\SO(N,5)}{\SO(N)\times \SO(5)}$. The superconformal theory
of the tensor multiplets (as world volume theory of $M$5 branes) was
constructed before in Ref. \cite{Claus:1997cq}. However, as mentioned
above, there are field equations but no action for this theory.

Thus, in conclusion, a superconformal construction exists for (2,0)
supergravity, for which there are field equations but no invariant
action. The (1,1) supergravity has no superconformal construction, due
to the absence of a suitable superconformal group. Therefore the
superconformal conjecture on $D=4$ is not invalidated by the current
knowledge about the superconformal models in $D=6$.

\

It is rather interesting to see how all facts known about these various
$N=4$ models seem to fall into place. Our conjecture, therefore, is that
new computations will continue to support the $N=4$ superconformal
symmetry of the model underlying pure $N=4$ supergravity.

\section{Discussion}

We have discussed here the pure $N=4$ Poincar{\'e} supergravity, which is a
gauge-fixed version of the corresponding $N=4$ superconformal theory,
the details of which, including the action in Eq. (3.16), are given by
de Roo in Ref. \cite{deRoo:1984gd}. Here we have explained briefly the
important details of the gauge fixing to  $N=4$ Poincar{\'e} supergravity at
the simple level of the bosonic part of the theory, as well as the role
of the conformal compensators, six vector multiplets with the wrong sign
kinetic terms. In particular, we have explicitly presented a triangular
gauge for the local $U(1)$ symmetry in which the superconformal model
\cite{deRoo:1984gd} becomes a pure $N=4$ supergravity model
\cite{Cremmer:1977tt}.

We argued that the $N=4$ superconformal action in Ref. \cite{deRoo:1984gd} is
unique and that the symmetry does not admit higher derivative actions.
The argument about  the uniqueness of the $N=4$ superconformal model is
based on the open gauge algebra of the $SU(2,2|4)$ superconformal
symmetry\footnote{Note that the algebra is closed on the Weyl multiplet,
therefore all local symmetry transformations are fixed. Meanwhile the
fact that the algebra is open on the $N=4$ Maxwell multiplet may be also
related to the need to use an infinite number of auxiliary fields to
close the algebra \cite{Siegel:1981dx}.}, which requires the  equations
of motion for the fermion partner of the compensator.  This allowed de
Roo in Ref. \cite{deRoo:1984gd} to reconstruct the action consistent with
the open algebra. Our argument about the uniqueness of the $N=4$
superconformal theory is related to the absence of  the higher
derivative superconformal invariants. Such invariants  require the
presence of negative conformal weight superfields, constructed from
conformal compensators,  which can be used in building new
superconformal invariants. Since in $N=4$  the only matter multiplets that
are available to serve as conformal compensators are vector multiplets
with the open algebra, they do not provide the negative weight
superfields which will allow to make
 the $N=4$ superconformal generalization of $
 \int d^4  x \,  \sqrt{-g} \,   \phi^{-4}  C^4$ counterterms, where $C$ is the Weyl tensor.
Therefore the $R^4$ UV divergences are forbidden by the $N=4$
superconformal symmetry of the un-gauge-fixed theory,  assuming that we
have all tools available for such a construction.

We have presented in the Appendix the detailed discussion of the
difficulties with the ``bottom up order by order attempts''  to
construct the corresponding higher derivative $N=4$ supergravity
invariants. One may try to start from the known on-shell $N=4$
superspace \cite{Brink:1979nt} candidate counterterms
\cite{Kallosh:1980fi,Bossard:2011tq} and  deform the classical
supersymmetry to reach the agreement with an exact deformation of
classical theory studied in the $N=2$ theory \cite{Chemissany:2012pf}.
The problem is the absence  of a clear guiding principle in the $N=4$
case. Therefore one can view the computation in Ref. \cite{Bern:2012cd}
as an indication that such a deformation may be indeed impossible since
we already have all tools available and they do not produce higher order
genuine supersymmetric invariants.

 We have analyzed the situation with $N=4$ local superconformal anomalies based on earlier detailed studies of consistent anomalies in $N=1$ superconformal theories in Ref. \cite{deWit:1985bn} and in
Refs. \cite{Schwimmer:2010za,Buchbinder:1988yu}. We argued that there is
no generalization of local superconformal $N=1$ anomalies to the $N=4$
case, the reason being the same as for counterterms. The anomaly
candidate requires us to use the superfield $\ln \phi$, the logarithm of
the compensator field, for constructing a consistent anomaly. But it is
not possible in $N=4$, for the same reason as the negative powers of
$\phi$ are not available as building blocks for superinvariants.

This observation provides  the simplest possible explanation of the
computation in Ref. \cite{Bern:2012cd} where $R^4$ UV divergence in
$N=4$, $L=3$ supergravity was found to cancel. Note that if this is the
true explanation, it would mean also that no other higher loop UV
divergences are predicted by the $N=4$ superconformal theory. Therefore
our conjecture is falsifiable; as soon as the UV properties of the
4-loop $N=4$ supergravity will be known, they will either confirm or
invalidate our conjecture.

The conjectured superconformal symmetry of $N=4$ supergravity supports
UV finiteness arguments for  $N\geq 4$ supergravities. For these models
the UV finiteness argument is associated with  the
Noether-Gaillard-Zumino deformed duality current conservation
\cite{Kallosh:2011dp} and  with local supersymmetry deformed by the
presence of the higher derivative superinvariant
\cite{Chemissany:2012pf}. Both arguments require the existence of the
Born-Infeld type deformation of extended supergravities
\cite{Carrasco:2011jv,Chemissany:2012pf}. In the particular case of
$N=4$ such a Born-Infeld type deformation is not possible according to
our current best understanding of superconformal symmetry, which is a
supporting argument for the UV finiteness of the $N> 4$ models. If the
$N=8$ Born-Infeld supergravity would be available, one would  be able to
derive the $N=4$ one by supersymmetry truncation, in conflict with
superconformal symmetry.

If our conjecture that the local superconformal symmetry explains the  3-loop UV finiteness in $N=4$ is confirmed by the
4-loop case, it will give us a hint that the models with
 superconformal symmetry without any dimensionful parameters may
serve as a basis for constructing a consistent quantum theory where
$M_{Pl}$ appears in the process of gauge fixing spontaneously broken
Weyl symmetry.

\section*{Acknowledgments}

 We are grateful to  Z. Bern, R. Roiban, M. Ro\v{c}ek, I. Tyutin for useful discussions and to A. Tseytlin  for a significant impact on the development of this investigation and to A. Linde and G. 't Hooft for the stimulating discussions of the potential role of the superconformal symmetry in physics.
 The work of S.F. is supported by the ERC Advanced Grant no. 226455, Supersymmetry, Quantum Gravity and Gauge Fields (SUPERFIELDS).
 The work of R.K. is supported  by SITP and NSF grant PHY-0756174 and by the Templeton grant ``Quantum Gravity Frontiers''.
 The work of A.V.P. is supported in part by the FWO - Vlaanderen, Project No.
G.0651.11, and in part by the European Science Foundation Holograv
Network.

 \appendix
 \section{Consistent deformation of $N=4$ supergravity?}

{\it Why at present there is a problem with a consistent  deformation of
pure $N=4$ supergravity with the $R^4$  term  and what has to be done to
solve it?}

Since $N=4$ pure supergravity is a gauge-fixed version of the
superconformal $N=4$ model, one can think about a deformation of the
theory to accommodate higher derivative actions either at the
superconformal level or at the level of the super Poincar{\'e} gauge-fixed
theory.

This appendix has a purpose to compare the model with $N=4$
superconformal symmetry with the one studied in $N=2$. Here we will
discuss the situation with $N=2$ supergravity at the super Poincar{\'e}
level, following \cite{Chemissany:2012pf}. In the $N=2$ case first the
genuine superconformal $N=2$ higher derivative action was provided
explicitly in Ref. \cite{deWit:2010za} in the general case and in the
simplest possible case corresponding to minimal pure $N=2$ supergravity
in Ref. \cite{Chemissany:2012pf}.

In $N=2$ we start with the superconformal action (\ref{faction}), where
it is known how to produce superinvariants depending on $S^{-2}$ and
$\bar S^{-2}$ chiral compensators. This action produces the $N=2$
superconformal version of the bosonic term given in (\ref{DSconf})
$\phi^{-4}(C_{....})^4$. As we explained in Sec. \ref{ss:higherderscN4},
no such superconformal invariant is available in the $N=4$ case.

However, we may try to continue bottom up and start with the already
gauge-fixed superconformal $N=4$ model, i. e., with $N=4$ supergravity
where at least the nonlinear on-shell supersymmetric $R^4$ counterterm
is available \cite{Kallosh:1980fi},\cite{Bossard:2011tq} based on the
on-shell superspace  construction \cite{Brink:1979nt}.
 In the absence of genuine local supersymmetry and in the absence of  auxiliary fields we can start from the classical action of $N=4$ supergravity and deform it by the known counterterm
 \be
S_1 ^{def}(\vf) = S_{0}(\vf) + \lambda \, S_{ct}(\vf)\,. \label{onc}
\ee
First we compute the variation of this action under undeformed local
transformation
\be
\delta_0 S_1^{def} =  {\delta S_0\over \delta \vf}
\delta_0 \vf+           \lambda
{\delta S_{ct}\over \delta \vf} \delta_0 \phi = \lambda {\delta
S_{ct}\over \delta \vf} \delta_0 \phi\,. \label{naive}
\ee
The first term in
(\ref{naive}) vanishes for generic field configurations according to
the definition of a local supersymmetry of the classical action.
  \be
 {\delta S_{0}\over \delta \vf} \delta_0 \phi =0\,.
 \ee

    The second term, the supersymmetry variation of the counterterm,
 vanishes only when the classical equations of motion are satisfied.
 Therefore the best we can say is that
\be
\delta_0 S_1^{def} =  \lambda
{\delta S_{ct}\over \delta \vf} \delta_0 \vf = \lambda {\delta
S_0\over \delta \vf} \delta X(\vf)
\label{naive1}
\ee
which generically
is not zero.\footnote{In general it is not zero but one can try to argue
that maybe it is actually zero, and in such case none of the problems
described in this appendix actually materialize. This is why this issue
was not given proper attention in the past. In Ref. \cite{Chemissany:2012pf}
it was shown that  the modification of the classical supersymmetry
transformation is necessary; the contribution from various  auxiliary
fields does not cancel for example in the nonlinear part of the
gravitino supersymmetry transformation.  In $N=4$ supergravity analogous
terms must be present to provide a consistent truncation from higher
supersymmetries to $N=2$.} In fact, the counterterm structure does not
allow an unambiguous extraction of what the $ \delta X(\vf)$ is since
the on-shell superspace construction \cite{Brink:1979nt} solves the
geometric Bianchi identities only under condition that
 \be
 {\delta S_{0}\over \delta \vf}  =0
 \label{onshell}
 \ee
 and therefore terms in the counterterms proportional to ${\delta S_{0}\over \delta \vf}$ are not unambiguously  defined.
 Since also the expression  ${\delta S_{0}\over \delta \vf}$ under local classical supersymmetry transforms via a linear combination of ${\delta S_{0}\over \delta \vf}$, none of these are
  directly available from the on-shell counterterms.

However, our recently acquired knowledge of the situation with genuine
 $N=2$ superinvariants where auxiliary fields are eliminated teaches us
that we have  to modify the symmetry transformations so that $ \delta
\vf = \delta_0 \vf + \lambda \delta_1 \vf$. We need to make the
following steps.
Assume that
we somehow succeed to generalize the known counterterm to the stage
where we can find $\delta X$ by performing the variation. We will call this generalization $ \hat S_{ct}$.
In such case we have
\be
{\delta \hat S_{ct}\over \delta
\vf} \delta_0 \vf = {\delta S_0\over \delta \vf} \delta X(\vf).
\label{naive2}
\ee
This reminds us the situation described in Sec. \ref{ss:higherderscN4} in Eq.
({\ref{onshellToine}) where the variation of the action under
supersymmetry is explicitly proportional to left-hand side of the Dirac
equation: if $\gamma^\mu \partial_\mu \chi=0$ the supersymmetry
variation of the action vanishes; otherwise it is proportional to
$\gamma^\mu \partial_\mu \chi$ and does not vanish.

Now we  get
\be
\delta S_1^{def} =  {\delta S_0\over \delta \vf} (\delta_0 \vf +\lambda  \delta_1 \vf)+   \lambda
{\delta \hat S_{ct}\over \delta \vf}    (\delta X+ \lambda \delta_1
\phi)\,.
\ee
Terms linear in $\lambda$ cancel if
\be
\delta_1 \vf= \delta X(\vf)\,.
\ee
If we find $\hat S_{ct}$ with computable $\delta X(\vf)$ in the $N=4$ case, we have identified $\delta_1 \vf$.
 We are then left with non-vanishing
\be
\lambda^2 {\delta \hat S_{ct}\over \delta \vf} \delta_1 \phi\,.
\ee
To cancel this one we have to find a next term in the action
\be
S^{def}_2(\vf)= S_{0}(\vf) + \lambda \, \hat S_{ct}(\vf)+ \lambda^2\,.
S_2(\vf)\,.
\ee
Assume that we can find the function $S_2(\vf)$ and the next order of deformation $\delta_2 \vf$ such that
\be
\delta S^{def}_2 = \lambda^2 \Big ( {\delta S_0\over \delta \vf}  \delta_2 \vf+
{\delta \hat S_{ct}\over \delta \vf}     \delta_1 \phi +{\delta S_2\over \delta \vf}\delta_0 \vf\Big )=0
\label{2}\ee
for generic configuration of $\phi$. This is an extremely strong condition: to find $S_2 (\vf  )  $ and $\delta_2 \vf( \vf    )$ such that the second term in (\ref{2}) will be compensated by
\be
{\delta S_0\over \delta \vf}  \delta_2 \vf
+{\delta S_2\over \delta \vf}\delta_0 \vf \,.
\ee
Assume this problem at the $\lambda^2$ level was solved.

Now we have the analogous problem at the $\lambda^3$ order when we take into account that we have suppressed the term
\be
\lambda^3 \Big ({\delta \hat S_{ct}\over \delta \vf}    \delta_2 \phi + {\delta S_2\over \delta \vf}\delta_1 \vf \Big
)\,.
\ee
We need to find $S_3$ and $\delta_3 \phi$.
Same for all higher order terms, we have to find new actions $S_n$ and extra symmetries $\delta_n
\vf$.

In $N=2$ we have a closed form answer,  the complete $\lambda$-independent local supersymmetry transformations  and the complete action in Eq. (\ref{faction}) which is linear in $\lambda$. Expanding around the classical solutions for auxiliary fields we reproduce a procedure analogous to one described here, since we can  we extract the values of $S_2 (\vf  )  $ and $\delta_2 \vf$,  ... , $S_n$ and $\delta_n \phi$  for any $n$ from the complete supersymmetric $N=2$ theory.

Meanwhile in $N=4$ the first step,  $S_{ct}$ $\Rightarrow$ $\hat S_{ct}$,  is not known, and  the infinite amount of next steps is also not known to exist and { \it there is no guiding principle}.  In a sense, step by step finding if this completion is possible is not much easier than computing the loop corrections.


\begin{thebibliography}{99}

%\cite{Ferrara:1977ij}
\bibitem{Ferrara:1977ij}
  S.~Ferrara, M.~Kaku, P.~K.~Townsend and P.~van Nieuwenhuizen,
  ``Gauging the graded conformal group with unitary internal symmetries,''
  Nucl.\ Phys.\ B {\bf 129}, 125 (1977).

  \bibitem{TheBook} D.Z. Freedman and A. Van  Proeyen, "Supergravity", Cambridge U.P.,
  2012.



  %\cite{Bergshoeff:1980sw}
\bibitem{Bergshoeff:1980sw}
  E.~Bergshoeff, M.~de Roo, J.~W.~van Holten, B.~de Wit and A.~Van Proeyen,
 ``Extended conformal supergravity and its applications,''
   in Superspace\& supergravity, ed. S.W. Hawking and M. Ro\v{c}ek (CUP, 1981) 237.
   %\cite{de Wit:1981gj}


%\cite{Bergshoeff:1980is}
\bibitem{Bergshoeff:1980is}
  E.~Bergshoeff, M.~de Roo and B.~de Wit,
``Extended conformal supergravity,''
  Nucl.\ Phys.\ B {\bf 182}, 173 (1981).
  %%CITATION = NUPHA,B182,173;%%


  %\cite{de Roo:1984gd}
\bibitem{deRoo:1984gd}
  M.~de Roo,
 ``Matter coupling in $N=4$ supergravity,''
  Nucl.\ Phys.\ B {\bf 255}, 515 (1985).
  %%CITATION = NUPHA,B255,515;%%

  %\cite{deRoo:1985jh}
\bibitem{deRoo:1985jh}
  M.~de Roo and P.~Wagemans,
  ``Gauge matter coupling in $N=4$ supergravity,''
  Nucl.\ Phys.\ B {\bf 262}, 644 (1985).
  %%CITATION = NUPHA,B262,644;%%

  %\cite{deWit:1978pd}
\bibitem{deWit:1978pd}
  B.~de Wit and S.~Ferrara,
  ``On higher order invariants in extended supergravity,''
  Phys.\ Lett.\ B {\bf 81}, 317 (1979).
  %%CITATION = PHLTA,B81,317;%%


%\cite{Tourkine:2012ip}
\bibitem{Tourkine:2012ip}
  P.~Tourkine and P.~Vanhove,
  ``An $R^4$ non-renormalisation theorem in $N=4$ supergravity,''
  Class.\ Quant.\ Grav.\  {\bf 29}, 115006 (2012)
  [arXiv:1202.3692 [hep-th]].
  %%CITATION = ARXIV:1202.3692;%%


  %\cite{Fischler:1979yk}
\bibitem{Fischler:1979yk}
  M.~Fischler,
``Finiteness calculations for O(4) through O(8) extended supergravity and O(4) supergravity coupled to selfdual O(4) matter,''
  Phys.\ Rev.\ D {\bf 20}, 396 (1979).
  %%CITATION = PHRVA,D20,396;%%
  %\cite{Bern:2012cd}
\bibitem{Bern:2012cd}
  Z.~Bern, S.~Davies, T.~Dennen and Y.~-t.~Huang,
  ``Absence of three-loop four-point divergences in $N=4$ Supergravity,''
  arXiv:1202.3423 [hep-th].
  %%CITATION = ARXIV:1202.3423;%%
  %\cite{Tourkine:2012ip}

   %\cite{Kallosh:2012ei}



\bibitem{Bern:2007xj}
  Z.~Bern, J.~J.~Carrasco, D.~Forde, H.~Ita and H.~Johansson,
  ``Unexpected cancellations in gravity theories,''
  Phys.\ Rev.\ D {\bf 77}, 025010 (2008)
  [arXiv:0707.1035 [hep-th]].
  %%CITATION = ARXIV:0707.1035;%%
\bibitem{mar}
  N.~Marcus,
  ``Composite anomalies in supergravity,''
  Phys.\ Lett.\ B {\bf 157}, 383 (1985);
  %%CITATION = PHLTA,B157,383;%%
  %\bi{dfg}
P.~di Vecchia, S.~Ferrara and L.~Girardello,
  ``Anomalies of hidden local chiral symmetries in sigma models and extended supergravities,''
  Phys.\ Lett.\ B {\bf 151}, 199 (1985).
  %%CITATION = PHLTA,B151,199;%%

\bibitem{Kallosh:2011dp}
  R.~Kallosh,
``$E_{7(7)}$ symmetry and finiteness of N=8 supergravity,''
  JHEP {\bf 1203}, 083 (2012)
  [arXiv:1103.4115 [hep-th]].
  %%CITATION = ARXIV:1103.4115;%%
\bibitem{Kallosh:2012ei}
  R.~Kallosh,
  ``On absence of 3-loop divergence in $N=4$ supergravity,''
  Phys.\ Rev.\ D {\bf 85}, 081702 (2012)
  [arXiv:1202.4690 [hep-th]].
  %%CITATION = ARXIV:1202.4690;%%
    %\cite{Deser:1970hs}
\bibitem{Deser:1970hs}
  S.~Deser,
  ``Scale invariance and gravitational coupling,''
    Annals Phys.\  {\bf 59}, 248 (1970);
  %%CITATION = APNYA,59,248;%%
%\cite{Anderson:1971dm}
%\bibitem{Anderson:1971dm}
  J.~L.~Anderson,  ``Scale invariance of the second kind and the brans-dicke scalar-tensor theory,''
    Phys.\ Rev.\ D {\bf 3}, 1689 (1971);
  %%CITATION = PHRVA,D3,1689;%%
  %\cite{Deser:1975sx}
 R.~E.~Kallosh,  ``On the renormalization problem of quantum gravity''
  Phys.\ Lett.\ B {\bf 55}, 321 (1975);
%\bibitem{Deser:1975sx}
  S.~Deser, M.~T.~Grisaru, P.~van Nieuwenhuizen and C.~C.~Wu,
  ``Scale dependence and the renormalization problem of quantum gravity,''
  Phys.\ Lett.\ B {\bf 58}, 355 (1975);
  T. Kugo and S. Uehara,  "Improved superconformal gauge conditions in
the $\mathcal{N}=1$ supergravity Yang-Mills matter system'', Nucl.
Phys. \textbf{B222}, 125 (1983);
  R.~Kallosh, L.~Kofman, A.~D.~Linde and A.~Van Proeyen,
   ``Superconformal symmetry, supergravity and cosmology,''
   Class.\ Quant.\ Grav.\  {\bf 17} (2000) 4269
   [Erratum-ibid.\  {\bf 21} (2004) 5017]
  [hep-th/0006179];
  %%CITATION = HEP-TH/0006179;%%
  %%CITATION = PHLTA,B58,355;%%
  %\cite{Bars:2010ae}
%\bibitem{Bars:2010ae}
  I.~Bars,
  ``Constraints on interacting scalars in 2T field theory and no scale models in 1T field theory,''
  Phys.\ Rev.\ D {\bf 82}, 125025 (2010)
  [arXiv:1008.1540 [hep-th]];
  %%CITATION = ARXIV:1008.1540;%%
%\cite{Ferrara:2010in}
%\bibitem{Ferrara:2010in}
  S.~Ferrara, R.~Kallosh, A.~Linde, A.~Marrani and A.~Van Proeyen,
  ``Superconformal symmetry, NMSSM, and inflation,''
  Phys.\ Rev.\ D {\bf 83}, 025008 (2011)
  [arXiv:1008.2942 [hep-th]].

   %\cite{Kallosh:1974yh}
\bibitem{Kallosh:1974yh}
  R.~E.~Kallosh,
 ``The renormalization in nonabelian gauge theories,''
  Nucl.\ Phys.\  B {\bf 78}, 293 (1974).
  %%CITATION = NUPHA,B78,293;%%
%\cite{vanNieuwenhuizen:1976vb}
%\bibitem{vanNieuwenhuizen:1976vb}
  P.~van Nieuwenhuizen and C.~C.~Wu,
 ``On integral relations for invariants constructed from three riemann tensors
and their applications in quantum gravity,''
  J.\ Math.\ Phys.\  {\bf 18}, 182 (1977).
  %%CITATION = JMAPA,18,182;%%



 %\cite{Goroff:1985th}
\bibitem{Goroff:1985th}
  M.~H.~Goroff and A.~Sagnotti,
``The ultraviolet behavior of Einstein gravity,''
  Nucl.\ Phys.\  B {\bf 266}, 709 (1986).
  %%CITATION = NUPHA,B266,709;%%
%\cite{vandeVen:1991gw}
%\bibitem{vandeVen:1991gw}
  A.~E.~M.~van de Ven,
``Two loop quantum gravity,''
  Nucl.\ Phys.\  B {\bf 378}, 309 (1992).
  %%CITATION = NUPHA,B378,309;%%


  %\cite{'tHooft:1974bx}
\bibitem{'tHooft:1974bx}
  G.~'t Hooft and M.~J.~G.~Veltman,
  ``One loop divergencies in the theory of gravitation,''
  Annales Poincare Phys.\ Theor.\ A {\bf 20}, 69 (1974).
  %%CITATION = AHPAA,A20,69;%%


   %\cite{Kallosh:1978wt}
\bibitem{Kallosh:1978wt}
  R.~E.~Kallosh, O.~V.~Tarasov and I.~V.~Tyutin,
 ``One loop finiteness of quantum gravity off mass shell,''
  Nucl.\ Phys.\ B {\bf 137}, 145 (1978).
  %%CITATION = NUPHA,B137,145;%%


  %\cite{Deser:1977nt}
\bibitem{Deser:1977nt}
  S.~Deser, J.~H.~Kay and K.~S.~Stelle,
``Renormalizability properties of supergravity,''
  Phys.\ Rev.\ Lett.\  {\bf 38}, 527 (1977).
  %%CITATION = PRLTA,38,527;%%
  %\cite{Deser:1978br}
%\bibitem{Deser:1978br}
  S.~Deser and J.~H.~Kay,
  ``Three loop counterterms for extended supergravity,''
  Phys.\ Lett.\ B {\bf 76}, 400 (1978);
  %%CITATION = PHLTA,B76,400;%%
  %\cite{Ferrara:1977mv}
%\bibitem{Ferrara:1977mv}
  S.~Ferrara and B.~Zumino,
  ``Structure of Conformal Supergravity,''
  Nucl.\ Phys.\ B {\bf 134}, 301 (1978).
  %%CITATION = NUPHA,B134,301;%%

   \bibitem{Toine} M. de Roo, J.W. van Holten, B. de Wit and A. Van Proeyen,
"Chiral superfields in $N=2$ supergravity, Nucl. Phys. B173 (1980)
175.

   %\cite{deWit:2010za}
\bibitem{deWit:2010za}
  B.~de Wit, S.~Katmadas and M.~van Zalk,
 ``New supersymmetric higher-derivative couplings: Full $N=2$ superspace does not count!,''
  JHEP {\bf 1101}, 007 (2011)
  [arXiv:1010.2150 [hep-th]].
  %%CITATION = ARXIV:1010.2150;%%


%\cite{Chemissany:2012pf}
\bibitem{Chemissany:2012pf}
  W.~Chemissany, S.~Ferrara, R.~Kallosh and C.~S.~Shahbazi,
  ``N=2 supergravity counterterms, off and on shell,''
  arXiv:1208.4801 [hep-th].
  %%CITATION = ARXIV:1208.4801;%%


   %\cite{Bossard:2011tq}
\bibitem{Bossard:2011tq}
  G.~Bossard, P.~S.~Howe, K.~S.~Stelle and P.~Vanhove,
  ``The vanishing volume of $D=4$ superspace,''
  Class.\ Quant.\ Grav.\  {\bf 28}, 215005 (2011)
  [arXiv:1105.6087 [hep-th]].
  %%CITATION = ARXIV:1105.6087;%%


   %\cite{Bern:2007hh}
\bibitem{Bern:2007hh}
  Z.~Bern, J.~J.~Carrasco, L.~J.~Dixon, H.~Johansson, D.~A.~Kosower and R.~Roiban,
  ``Three-loop superfiniteness of $N=8$ supergravity,''
  Phys.\ Rev.\ Lett.\  {\bf 98}, 161303 (2007)
  [hep-th/0702112].
  %%CITATION = HEP-TH/0702112;%%
Z.~Bern, J.~J.~M.~Carrasco, L.~J.~Dixon, H.~Johansson and R.~Roiban,
  ``Simplifying multiloop integrands and ultraviolet divergences of gauge theory and gravity amplitudes,''
  Phys.\ Rev.\ D {\bf 85}, 105014 (2012)
  [arXiv:1201.5366 [hep-th]].
  %%CITATION = ARXIV:1201.5366;%%

Z.~Bern, J.~J.~Carrasco, L.~J.~Dixon, H.~Johansson and R.~Roiban,
  ``The ultraviolet behavior of $N=8$ supergravity at four loops,''
  Phys.\ Rev.\ Lett.\  {\bf 103}, 081301 (2009)
  [arXiv:0905.2326 [hep-th]].
  %%CITATION = ARXIV:0905.2326;%%





 %\cite{Cremmer:1977tt}
\bibitem{Cremmer:1977tt}
  E.~Cremmer, J.~Scherk and S.~Ferrara,
  ``SU(4) invariant supergravity theory,''
  Phys.\ Lett.\ B {\bf 74}, 61 (1978).
  %%CITATION = PHLTA,B74,61;%%


\bibitem{Das:1977uy}
  A.~K.~Das,
  ``SO(4) invariant extended supergravity,''
  Phys.\ Rev.\ D {\bf 15}, 2805 (1977).
  %%CITATION = PHRVA,D15,2805;%%
%\cite{Cremmer:1977tc}
%\bibitem{Cremmer:1977tc}
  E.~Cremmer and J.~Scherk,
  ``Algebraic simplifications in supergravity theories,''
  Nucl.\ Phys.\ B {\bf 127}, 259 (1977).
  %%CITATION = NUPHA,B127,259;%%



%\cite{Cremmer:1979up}
\bibitem{Cremmer:1979up}
  E.~Cremmer and B.~Julia,
  ``The SO(8) supergravity,''
  Nucl.\ Phys.\ B {\bf 159}, 141 (1979).
  %%CITATION = NUPHA,B159,141;%%


  %\cite{deWit:1985bn}
\bibitem{deWit:1985bn}
  B.~de Wit and M.~T.~Grisaru,
  ``Compensating fields and anomalies,''
Edited by I. A. Batalin et al.: Quantum field theory and quantum
statistics, Adam Hilger, Bristol Vol. 2, 1987.

  \bibitem{Schwimmer:2010za}
  A.~Schwimmer and S.~Theisen,
 ``Spontaneous breaking of conformal invariance and trace anomaly matching,''
  Nucl.\ Phys.\ B {\bf 847}, 590 (2011)
  [arXiv:1011.0696 [hep-th]].
  %%CITATION = ARXIV:1011.0696;%%


   %\cite{Buchbinder:1988yu}
\bibitem{Buchbinder:1988yu}
  I.~L.~Buchbinder and S.~M.~Kuzenko,
 ``Nonlocal action for supertrace anomalies in superspace of $N=1$ supergravity,''
  Phys.\ Lett.\ B {\bf 202}, 233 (1988).
  %%CITATION = PHLTA,B202,233;%%

   %\cite{Wess:1971yu}
\bibitem{Wess:1971yu}
  J.~Wess and B.~Zumino,
 ``Consequences of anomalous Ward identities,''
  Phys.\ Lett.\ B {\bf 37}, 95 (1971).
  %%CITATION = PHLTA,B37,95;%%



   %\cite{Kallosh:1972ap}
\bibitem{Kallosh:1972ap}
  R.~E.~Kallosh and I.~V.~Tyutin,
  ``The equivalence theorem and gauge invariance in renormalizable theories,''
  Yad.\ Fiz.\  {\bf 17}, 190 (1973)
  [Sov.\ J.\ Nucl.\ Phys.\  {\bf 17}, 98 (1973)].
  %%CITATION = YAFIA,17,190;%%


\bibitem{Arkady} I.~L.~Buchbinder, N.~G.~Pletnev and A.~A.~Tseytlin,
``'Induced' $N=4$ conformal supergravity,''
  Phys.\ Lett.\ B {\bf 717}, 274 (2012)
  [arXiv:1209.0416 [hep-th]].
  %%CITATION = ARXIV:1209.0416;%%
    %\cite{Fradkin:1985am}
\bibitem{Fradkin:1985am}
  E.~S.~Fradkin and A.~A.~Tseytlin,
  ``Conformal supergravity,''
  Phys.\ Rept.\  {\bf 119}, 233 (1985).
  %%CITATION = PRPLC,119,233;%%

\bibitem{Pasti:1996vs}
  P.~Pasti, D.~P.~Sorokin and M.~Tonin,
  ``On Lorentz invariant actions for chiral $p$ forms,''
  Phys.\ Rev.\ D {\bf 55} (1997) 6292
\href{http://arxiv.org/abs/hep-th/9611100}{{\tt hep-th/9611100}}.
  %%CITATION = HEP-TH/9611100;%%

  %\cite{Bergshoeff:2001pv}
\bibitem{Bergshoeff:2001pv}
  E.~Bergshoeff, R.~Kallosh, T.~Ortin, D.~Roest and A.~Van Proeyen,
  ``New formulations of $D = 10$ supersymmetry and D8 - O8 domain walls,''
  Class.\ Quant.\ Grav.\  {\bf 18}, 3359 (2001)
  [hep-th/0103233].
  %%CITATION = HEP-TH/0103233;%%
\bibitem{Claus:1997cq}
  P.~Claus, R.~Kallosh and A.~Van Proeyen,
  ``M 5-brane and superconformal (0,2) tensor multiplet in six-dimensions,''
  Nucl.\ Phys.\ B {\bf 518} (1998) 117
\href{http://arxiv.org/abs/hep-th/9711161}{{\tt hep-th/9711161}}.].
  %%CITATION = HEP-TH/9711161;%%}
  \bibitem{Giani:1984dw}
  F.~Giani, M.~Pernici and P.~van Nieuwenhuizen,
  ``Gauged $N=4$ $d = 6$ supergravity,''
  Phys.\ Rev.\ D {\bf 30} (1984) 1680.
  %%CITATION = PHRVA,D30,1680;%%
\bibitem{Romans:1985tw}
  L.~J.~Romans,
  ``The F(4) gauged supergravity in six-dimensions,''
  Nucl.\ Phys.\ B {\bf 269} (1986) 691.
  %%CITATION = NUPHA,B269,691;%%
\bibitem{realLieSA}
V.G. Kac, ``Lie superalgebras,'' Adv. Math. {\bf 26} (1977) 8;
\\ M. Parker,
``Classification of real simple Lie superalgebras of classical type,'' J.
Math. Phys. {\bf 21} (1980) 689.
\bibitem{VanProeyen:1999ni}
A.~Van~Proeyen, ``Tools for supersymmetry,'' Ann. Univ.
 Craiova, Phys. AUC 9 (part~I) (1999) 1--48,
\href{http://arxiv.org/abs/hep-th/9910030}{{\tt hep-th/9910030}}
%%CITATION = HEP-TH 9910030;%%,
%\href{itf.fys.kuleuven.be/supergravity/pdf/ExtraAppB.pdf}{itf.fys.kuleuven.be/supergravity/pdf/ExtraAppB.pdf}.

\bibitem{nahm}
W. Nahm, ``Supersymmetries and their representations,'' Nucl. Phys. {\bf
B135} (1978) 149.  \bibitem{Riccioni:1997np}
  F.~Riccioni,
  ``Tensor multiplets in six-dimensional (2,0) supergravity,''
  Phys.\ Lett.\ B {\bf 422} (1998) 126
\href{http://arxiv.org/abs/hep-th/9712176}{{\tt hep-th/9712176}}.
  %%CITATION = HEP-TH/9712176;%%
\bibitem{Bergshoeff:1999db}
  E.~Bergshoeff, E.~Sezgin and A.~Van Proeyen,
  ``(2,0) tensor multiplets and conformal supergravity in $D = 6$,''
  Class.\ Quant.\ Grav.\  {\bf 16} (1999) 3193
\href{http://arxiv.org/abs/hep-th/9904085}{{\tt hep-th/9904085}}.
  %%CITATION = HEP-TH/9904085;%%

 %\cite{Siegel:1981dx}
\bibitem{Siegel:1981dx}
  W.~Siegel and M.~Ro\v{c}ek,
  ``On off-shell supermultiplets,''
  Phys.\ Lett.\ B {\bf 105}, 275 (1981).
  %%CITATION = PHLTA,B105,275;%%

   %\cite{Brink:1979nt}
\bibitem{Brink:1979nt}
L.~Brink and P.~S.~Howe,
``The ${\cal{N}}=8$ supergravity in superspace,''
Phys.\ Lett.\  B {\bf 88}, 268 (1979).
%%CITATION = PHLTA,B88,268;%%
  %\cite{Howe:1981gz}
%\bibitem{Howe:1981gz}
P.~S.~Howe,
``Supergravity in superspace,''
Nucl.\ Phys.\ B {\bf 199}, 309 (1982).
%%CITATION = NUPHA,B199,309;%%


  %\cite{Kallosh:1980fi}
\bibitem{Kallosh:1980fi}
R.~E.~Kallosh,
``Counterterms in extended supergravities,''
Phys.\ Lett.\  B {\bf 99} (1981) 122;
%%CITATION = PHLTA,B99,122;%%
P.~S.~Howe and U.~Lindstrom,
``Higher order invariants in extended supergravity,''
Nucl.\ Phys.\  B {\bf 181}, 487 (1981).
%%CITATION = NUPHA,B181,487;%%
P.~S.~Howe, K.~S.~Stelle and P.~K.~Townsend,
``Superactions,''
Nucl.\ Phys.\  B {\bf 191}, 445 (1981).
%%CITATION = NUPHA,B191,445;%%




  %\cite{Carrasco:2011jv}
\bibitem{Carrasco:2011jv}
  J.~J.~M.~Carrasco, R.~Kallosh and R.~Roiban,
  ``Covariant procedures for perturbative non-linear deformation of duality-invariant theories,''
  Phys.\ Rev.\ D {\bf 85}, 025007 (2012)
  [arXiv:1108.4390 [hep-th]].
  %%CITATION = ARXIV:1108.4390;%%
  %\cite{Bossard:2011ij}
%\bibitem{Bossard:2011ij}
  G.~Bossard and H.~Nicolai,
  ``Counterterms vs. dualities,''
  JHEP {\bf 1108}, 074 (2011)
  [arXiv:1105.1273 [hep-th]].
  %%CITATION = ARXIV:1105.1273;%%
  %\cite{Chemissany:2011yv}
%\bibitem{Chemissany:2011yv}
  W.~Chemissany, R.~Kallosh and T.~Ort{\'\i}n,
  ``Born-Infeld with higher derivatives,''
  Phys.\ Rev.\ D {\bf 85}, 046002 (2012)
  [arXiv:1112.0332 [hep-th]].
  %%CITATION = ARXIV:1112.0332;%%
  %\cite{Broedel:2012gf}
%\bibitem{Broedel:2012gf}
  J.~Broedel, J.~J.~M.~Carrasco, S.~Ferrara, R.~Kallosh and R.~Roiban,
  ``N=2 supersymmetry and U(1)-duality,''
  Phys.\ Rev.\ D {\bf 85}, 125036 (2012)
  [arXiv:1202.0014 [hep-th]].
  %%CITATION = ARXIV:1202.0014;%%
%\cite{Kuzenko:2012ht}
%\bibitem{Kuzenko:2012ht}
  S.~M.~Kuzenko,
  ``Nonlinear self-duality in $N = 2$ supergravity,''
  JHEP {\bf 1206}, 012 (2012)
  [arXiv:1202.0126 [hep-th]].
  %%CITATION = ARXIV:1202.0126;%%

   %\cite{Kallosh:2012yy}
%\bibitem{Kallosh:2012yy}
  R.~Kallosh and T.~Ort{\'\i}n,
  ``New $E_{7(7)}$ invariants and amplitudes,''
  JHEP {\bf 1209}, 137 (2012)
  [arXiv:1205.4437 [hep-th]].
  %%CITATION = ARXIV:1205.4437;%%




\end{thebibliography}
\end{document}